\documentclass[aps, twocolumn, showpacs, letterpaper]{revtex4}

\pdfoutput=1

\usepackage{amsmath}
\usepackage{amssymb}
\usepackage{graphicx}
\usepackage{xspace}
\usepackage{upgreek}
\usepackage{accents}

\newcommand{\eg}{{e.g.,\/}\xspace}
\newcommand{\ie}{{i.e.,\/}\xspace}

\newcommand{\eq}[1]{(\ref{#1})}
\newcommand{\Eq}[1]{Eq.~(\ref{#1})}
\newcommand{\Eqs}[1]{Eqs.~(\ref{#1})} 
\newcommand{\Fig}[1]{Fig.~\ref{#1}} 
\newcommand{\Ref}[1]{Ref.~\cite{#1}}
\newcommand{\Refs}[1]{Refs.~\cite{#1}}
\newcommand{\Sec}[1]{Sec.~\ref{#1}}
\newcommand{\App}[1]{Appendix~\ref{#1}}

\newcommand{\mc}[1]{\mathcal{#1}}
\newcommand{\mcc}[1]{\mathfrak{#1}}
\newcommand{\msf}[1]{\mathsf{#1}}

\newcommand{\favr}[1]{\langle #1 \rangle}

\newcommand{\pd}{\partial}

\newcommand{\kpt}[1]{{\kern #1 pt}}

\newcommand{\ds}{\displaystyle}

\begin{document}

\title{Adiabatic nonlinear waves with trapped particles: II. Wave dispersion}
\author{I.~Y. Dodin and N.~J. Fisch}
\affiliation{Department of Astrophysical Sciences, Princeton University, Princeton, New Jersey 08544, USA}
\date{\today}


% 52.35.-g -- Waves, oscillations, and instabilities in plasmas and intense beams
% 52.35.Mw -- Nonlinear phenomena: waves, wave propagation, and other interactions 
%             (including parametric effects, mode coupling, ponderomotive effects, etc.) 
% 52.25.-b -- Plasma properties
% 45.20.Jj -- Lagrangian and Hamiltonian mechanics 

\begin{abstract}
A general nonlinear dispersion relation is derived in a nondifferential form for an adiabatic sinusoidal Langmuir wave in collisionless plasma, allowing for an arbitrary distribution of trapped electrons. The linear dielectric function is generalized, and the nonlinear kinetic frequency shift $\omega_{\rm NL}$ is found analytically as a function of the wave amplitude $a$. Smooth distributions yield $\omega_{\rm NL} \propto \sqrt{a}$, as usual. However, beam-like distributions of trapped electrons result in different power laws, or even a logarithmic nonlinearity, which are derived as asymptotic limits of the same dispersion relation. Such beams are formed whenever the phase velocity changes, because the trapped distribution is in autoresonance and thus evolves differently from the passing distribution. Hence, even adiabatic $\omega_{\rm NL}(a)$ is generally nonlocal. 
\end{abstract}

\maketitle
\bibliographystyle{brief}

%%%%%%%%%%%%%%%%%%%%%%%%%%%%%%%%%%%%%%%%%%%%%%%%%%%%%%%%%%%%
\section{Introduction}

Within the geometrical-optics approximation, adiabatic waves in collisionless plasma are described conveniently within the average-Lagrangian formalism originally proposed by Whitham \cite{ref:whitham65, book:whitham}. In our \Ref{tex:myacti}, further called Paper I, this formalism was restated to also accommodate effects of particles trapped in wave troughs. Specifically, the wave Lagrangian density $\mcc{L}$ was derived explicitly under the assumption that the number of trapped particles within each wavelength is a fixed independent parameter of the problem. Hence the wave nonlinear dispersion relation (NDR) can be expressed in a nondifferential form which is more easily tractable than solutions obtained straightforwardly from integrating the Vlasov-Maxwell system (see, \eg \Refs{ref:schamel00, ref:krasovskii89, ref:krasovskii95, ref:benisti07, ref:matveev09, ref:bohm49}) or those inferred from the ``multiwater bag'' formulation \cite{ref:khain07}. Namely, the general NDR for an adiabatic wave reads~as (Paper~I)
\begin{gather}\label{eq:ndr0}
\pd_a \mcc{L} = 0,
\end{gather}
$a$ being the wave amplitude; thus, analyzing the function $\pd_a \mcc{L}$ would be sufficient to infer the wave dispersion.

Here, we perform a systematic study of kinetic nonlinearities that \Eq{eq:ndr0} yields for one-dimensional sinusoidal electrostatic waves with trapped particles, extending our brief letter~\cite{my:bgk} and providing important technical details missing in that work. Specifically, a general NDR is derived in a nondifferential form for an adiabatic sinusoidal Langmuir wave in collisionless plasma, allowing for an arbitrary distribution of trapped electrons [\Eq{eq:drpw}]. The linear dielectric function is generalized, and the nonlinear kinetic frequency shift $\omega_{\rm NL}$ is found analytically as a function of the wave amplitude $a$. Smooth distributions yield $\omega_{\rm NL} \propto \sqrt{a}$, as usual. However, beam-like distributions of trapped electrons result in different power laws, or even a logarithmic nonlinearity, which are derived as asymptotic limits of the same dispersion relation. Such beams are formed whenever the phase velocity changes, because the trapped distribution is in autoresonance and thus evolves differently from the passing distribution; hence, even adiabatic $\omega_{\rm NL}(a)$ is generally nonlocal. These results are also used in Paper~III \cite{tex:myactiii}, where evolution of waves with trapped particles is studied.

The work is organized as follows. In \Sec{sec:master}, the general NDR is derived, connecting the wave frequency $\omega$ with the wave number $k$ for a given $a$ and particle distribution; also, the properties of the NDR kernel are explored. In \Sec{sec:smooth}, the general NDR is given an asymptotic representation for distributions smooth near the resonance; in particular, the linear dielectric function is generalized. In \Sec{sec:beam}, the effect of particle beams on NDR is studied. In \Sec{sec:discuss}, implications of these results are discussed; specifically, $\omega_{\rm NL}(a)$ is predicted to be a nonlocal function, and the friction drag on trapped particles is predicted to affect the wave frequency sweeping in collisional plasmas. In \Sec{sec:concl}, we summarize our main results. Some auxiliary formulas are also presented in the appendix.

%%%%%%%%%%%%%%%%%%%%%%%%%%%%%%%%%%%%%%%%%%%%%%%%%%%%%%%%%%%%
\section{General dispersion relation}
\label{sec:master}

%-----------------------------------------------------------
\subsection{Wave Lagrangian}
\label{sec:wlagr}

From Paper~I, it follows that the Lagrangian density for a one-dimensional adiabatic nonlinear electrostatic electron mode in collisionless plasma reads~as
\begin{gather}\label{eq:lmain}
\mcc{L} = \bar{\mcc{L}}_{\rm em} + \frac{\favr{\tilde{E}^2}}{8\pi}- n^{(p)}\favr{\mc{H}^{(p)}}_{f_p} - n^{(t)}\favr{\mc{H}^{(t)}}_{f_t},
\end{gather}
assuming that the ion contribution is negligible. Here, $\bar{\mcc{L}}_{\rm em}$ describes quasistatic fields (if any), the second term is the time-averaged energy density of the wave electric field $\tilde{E}$, $n^{(p)}$ and $n^{(t)}$ are the densities of passing and trapped electrons, respectively, and $\mc{H}^{(p)}$ and $\mc{H}^{(t)}$ are the oscillation-center (OC) energies averaged over the corresponding distributions $f_p$ and $f_t$ of the particle canonical momenta. Specifically,  
\begin{gather}
\mc{H}^{(p)} = \mc{E} + Pu - mu^2/2, \label{eq:hp}\\
\mc{H}^{(t)} = \mc{E} - mu^2/2, \label{eq:ht}
\end{gather}
where $m$ is the electron mass, and $\mc{E}(J, a)$ is the electron energy in the frame $\hat{K}$ moving with respect to the laboratory frame $K$ at the wave phase velocity $u = \omega/k$. Also, $J$ is the trapped-particle canonical momentum, or the canonical action in $\hat{K}$, and
\begin{gather} \label{eq:canP}
P = mu \pm kJ 
\end{gather}
is the passing-particle canonical momentum, where the plus sign is chosen for particles traveling with velocities larger than $u$ and the minus sign is chosen otherwise.

Combining \Eqs{eq:lmain}-\eq{eq:ht}, one gets
\begin{gather}\notag
\mcc{L} = \bar{\mcc{L}}_{\rm em} + \frac{nmu^2}{2} - n_p^{(p)}\favr{P}_{f_p}u + \frac{\favr{\tilde{E}^2}}{8\pi} - n\favr{\mc{E}}_{f},
\end{gather}
where we introduced the electron total density $n = n^{(p)} + n^{(t)}$, and the energy $\mc{E}$ is now being averaged over all particles. By definition (Paper~I), the particle densities, $\omega$, and $k$ are arguments of $\mcc{L}$ that are independent of $a$. [It is only through \textit{solutions} of \Eq{eq:ndr0} that they will become interconnected.] Hence, \Eq{eq:ndr0} yields the general NDR in the~form
\begin{gather}\label{eq:ndr1}
\pd_a \favr{\tilde{E}^2}/(8\pi) - n\, \pd_a \favr{\mc{E}}_{f} = 0,
\end{gather}
where $\pd_a$ is taken at constant $(\omega,k)$ and also at fixed canonical momenta of individual particles. Specifically, fixed at differentiation can be $J$, not only for trapped particles but also for passing ones, due to \Eq{eq:canP}.

%-----------------------------------------------------------
\subsection{Sinusoidal-wave approximation}

Equation \eq{eq:ndr1} describes the dispersion of one-dimensional electron Langmuir waves completely, accounting simultaneously for linear and nonlinear dispersive effects, both fluid and kinetic. In principle, each of those effects can then be assessed from the general NDR by substituting the appropriate $\mc{E}(J, a)$ \cite{foot:anharm}. Yet, we will address only one specific regime here, as an illustration of how our Lagrangian formalism can be useful for finding NDRs explicitly; namely, we will assume that the wave can be considered monochromatic. The conditions under which this assumption is justified as a precise asymptotic limit at small amplitudes are discussed in detail in \Refs{ref:winjum07, ref:krasovsky94} (see also \Refs{ref:rose01, ref:krasovsky07}) and will not be repeated here. Instead, our purpose will be to show that, \textit{once} the sinusoidal-wave approximation has been adopted (like in traditional calculations), our Lagrangian formalism predicts new effects and also yields simultaneously various types of kinetic nonlinearities that were previously known from different contexts.

\begin{figure}
\centering
\includegraphics[width=.48\textwidth]{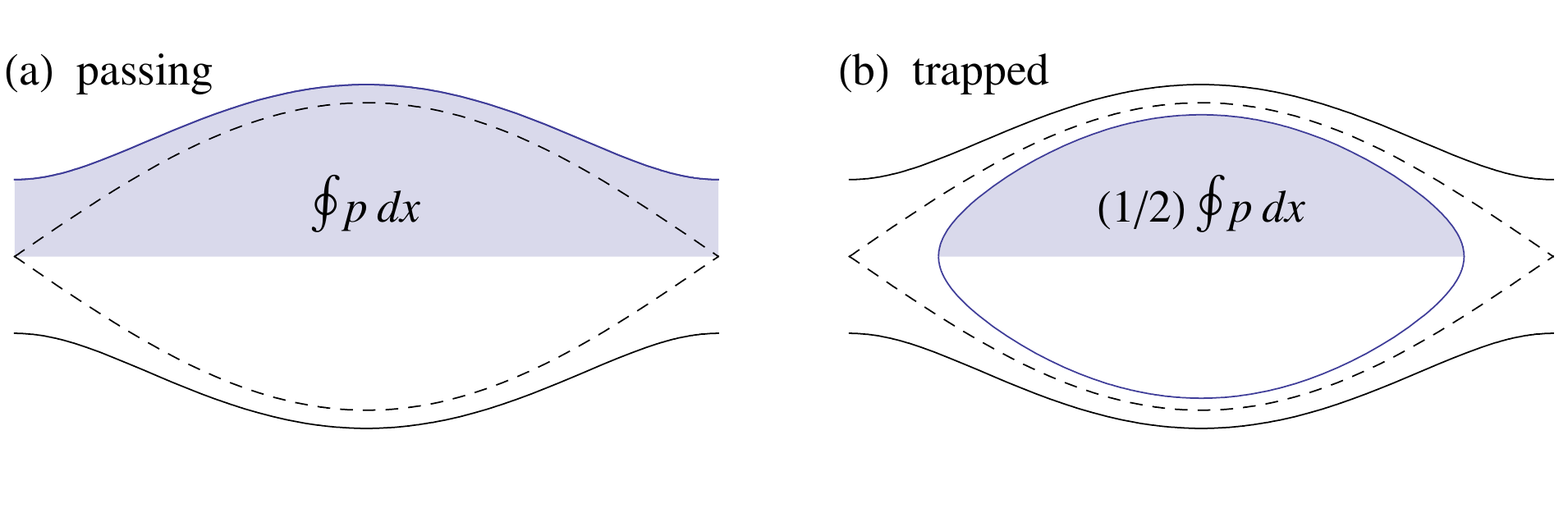}
\caption{Schematic of single-particle trajectories in phase space $(x,p)$, illustrating the definition of $2\pi J$ (shaded area): (a) for a passing particle, (b) for a trapped particle. For $J$ to be continuous at the separatrix (dashed), with the passing-particle action defined as $2\pi J = \oint p\,dx$, for trapped particles one must use the definition $2\pi J = (1/2) \oint p\,dx$. This area is encircled by a particle within half of the bounce period; thus, the corresponding canonical frequency $\Omega$ equals \textit{twice} the bounce frequency.}
\label{fig:J}
\end{figure}

In a sinusoidal wave, the oscillations of both passing and trapped electrons in $\hat{K}$ satisfy
\begin{gather}\label{eq:hamr}
p^2/(2m) + e\tilde{\varphi}_0 \cos (kx) = \mc{E},
\end{gather}
where $x$ and $p$ are the particle coordinate and momentum, and $e$ is the electron charge. (For clarity, we will assume $e\tilde{\varphi}_0 > 0$ and $k > 0$. Also, the large-scale electrostatic potential $\bar{\varphi}$ is omitted here because it does not affect the wave local properties, albeit it may affect the dynamics of the wave (Paper~III) and of the particle distribution \cite{ref:lindberg07}.) The canonical action $J$ is hence introduced as follows. For passing particles we will use the definition $J = (2\pi)^{-1} \oint p\,dx$, and thus $J = |p|/k$ for large $p$. For trapped particles, however, we will use $J = (4\pi)^{-1} \oint p\,dx$, so that $J$ is continuous across the separatrix (\Fig{fig:J}). From \Eq{eq:hamr}, one can write then
\begin{gather}\label{eq:J}
 J = \sqrt{a}\,\hat{J}j.
\end{gather}
Here we introduced the wave amplitude as $a = {e\tilde{\varphi}_0k^2/(m\omega^2) > 0}$; also, $\hat{J} = m\omega/k^2$, and the dimensionless action $j$ is a continuous function of the normalized energy $r = (\mc{E} + e\tilde{\varphi}_0)/(2e\tilde{\varphi}_0)$,
\begin{gather}\label{eq:jgeneral}
 j = j(r).
\end{gather}
Specifically, $j = 0$ for a particle resting at the bottom of the wave trough ($r = 0$), with the corresponding value at the separatrix ($r = 1$) being $j_* = 4/\pi$ (\App{app:func}). 

Hence, one can express $\mc{E}$ as
\begin{gather}
 \mc{E}(J, a) = [2ar(j) - a]\hat{J}\omega,
\end{gather}
with $j(J, a)$ taken from \Eq{eq:J}. Then \Eq{eq:ndr1} reads~as
\begin{gather}\label{eq:drpw}
\omega^2 = \omega_p^2\,\frac{2}{a}\int^\infty_0 G F(J)\,dJ,
\end{gather}
where $\omega_p^2 = 4\pi n e^2/m$ is the electron plasma frequency squared, $F(J)$ is the action distribution normalized such that $\int^\infty_0 F(J)\,dJ = 1$, and
\begin{gather}\label{eq:g0}
G = \frac{\pd_a\mc{E}(J, a)}{mu^2}.
\end{gather}

Equation \eq{eq:drpw} is a master equation, which accounts for \textit{all} dispersive effects simultaneously, both linear and nonlinear (to the extent that the sinusoidal-wave approximation applies \cite{foot:fluid}). In particular, it shows that the contribution to $\omega^2$ of particles with given $J$ is determined solely by the kernel $G$, thus acting as the weight function. We study the properties of this function below.

%----------------------------------------------------------
\subsection{Weight function \textit{\textbf{G}}}

From \Eq{eq:g0}, one gets
\begin{gather}
G = [\pd_a (2ar - a)]_J =  2r - 1 + 2a\, (\pd_a r)_J,
\end{gather}
where the partial derivative is taken at fixed $J$ (as denoted by the subindex). The latter term is found from
\begin{gather}
(\pd_a r)_J = - \frac{(\pd_a J)_r}{(\pd_r J)_a} = - \frac{j}{2a j'},
\end{gather}
where we substituted \Eq{eq:J}. Thus, $G = 2r - 1 - j/j'$, where the right-hand side here is a function of $r$ only, or a function of $j$ only [due to \Eq{eq:jgeneral}]. We will introduce these two representations separately, with 
\begin{gather}\label{eq:grgeneral}
g(r) = 2r - 1 - j(r)/j'(r)
\end{gather}
(\App{app:func}) serving as an auxiliary function for
\begin{gather}
G(j) = g(r(j)).
\end{gather}
Hence, the properties of $G(j)$ are understood as follows.

\begin{figure}
\centering
\includegraphics[width=.48\textwidth]{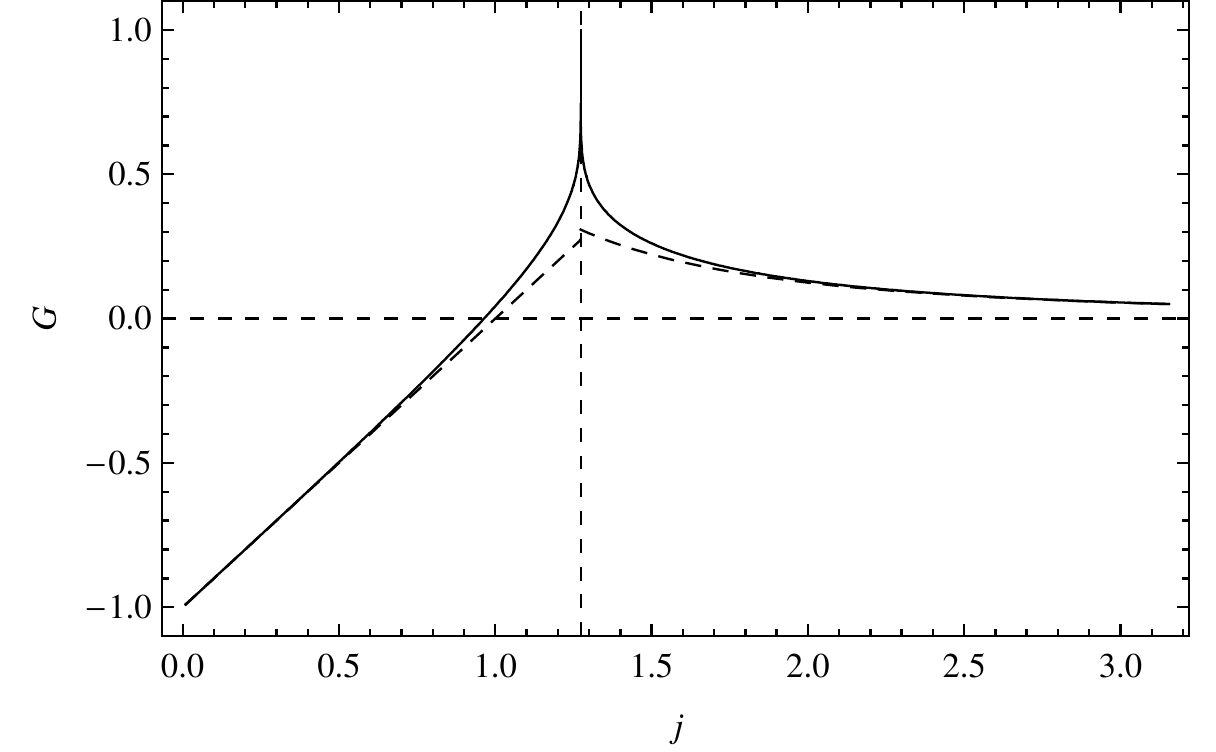}
\caption{Weight function $G(j)$. Dashed are approximate solutions given by \Eq{eq:gasym}, the asymptote, and $j = j_*$.}
\label{fig:g}
\end{figure}

First, let us consider asymptotics of $G(j)$, starting with $j \ll 1$. As seen from \Eq{eq:grgeneral}, $G(0) = -1$, which also flows from the fact that $\mc{E} \approx - e\tilde{\varphi}_0$ for particles close to the bottom of the wave trough. The next-order term can be inferred from $\mc{E} \approx - e\tilde{\varphi}_0 + J \Omega$, which is due to oscillations of deeply trapped particles being approximately linear. In particular, one can take $\Omega \approx 2\omega_E$ (\Fig{fig:J}), where $\omega_E = (e k^2\tilde{\varphi}_0/m)^{1/2} = \omega \sqrt{a}$ is the linear bounce frequency \cite{book:stix}. Hence, \Eq{eq:g0} yields $G \approx - 1 + j$.

The asymptotic behavior of $G(j)$ at $j \gg 1$ is understood by taking $\mc{E} \approx p^2/(2m) + \Phi$, where $\Phi = e^2m \tilde{\varphi}_0^2/(4p^2)$ is the ponderomotive energy in the dipole approximation (Paper~I) for a zero-frequency wave [\Eq{eq:hamr}]. Remember that $|p| \approx Jk$ in this case, so $(\pd_a \mc{E})_J = 2\Phi/a$. Thus,
\begin{gather}\label{eq:glargej}
G(j \gg 1) \approx 1/(2j^2),
\end{gather} 
and higher-order corrections can be found similarly.

While negative at small~$j$, $G(j)$ is positive at large~$j$. (This means that deeply trapped particles reduce $\omega^2$, whereas those near the separatrix and untrapped ones increase $\omega^2$; cf. \Ref{ref:dewar72b, ref:ikezi78}).) Now let us consider its behavior near the separatrix, which is where the canonical frequency $\Omega$ approaches zero. Since $\mc{E}$ equals the OC energy in $\hat{K}$ (Paper~I), one has
\begin{gather}\label{eq:Omega}
\Omega = \pd_J \mc{E} = \omega\, \pd_j (2ar-a) = 2a\omega/j'.
\end{gather}
Therefore $j'(r = 1)$ is infinite, and \Eq{eq:grgeneral} yields that $G$ be continuous (albeit nonanalytic), with $G(j_*) = 1$. Hence, we summarize our results as follows:
\begin{gather}\label{eq:gasym}
G(j) = \left\{ 
\begin{array}{ll}
\ds -1 + j + \ldots, & \quad j \ll 1,\\[5pt]
\ds \,\,\,1, & \quad j = 1, \\[3pt]
\ds \frac{1}{2j^2} + \frac{5}{16j^6}+ \ldots, & \quad j \gg 1.
\end{array} 
\right.
\end{gather}
These are matched (\Fig{fig:g}) by direct calculations in which an explicit expression for $j(r)$ is substituted into \Eq{eq:grgeneral}. Also, manipulating that and related formulas given in \App{app:func} shows that the NDR arising from the waterbag formulation (\Ref{ref:khain07}, with zero driving field) can, in principle \cite{foot:amend}, be reduced to our \Eq{eq:drpw}.

\subsection{Small-amplitude cold-plasma limit}

It is instructive to consider, as an example, the cold-plasma case in the limit of small amplitudes. Then, all particles are passing, and $J = |p|/k$,~so (Paper~I)
\begin{gather}\label{eq:vlin}
v^{\pm} = u \pm kJ/m,
\end{gather}
where the sign index denotes $\text{sgn}\,(v-u)$. Yet $v^+$ are not present then, and $v^{-} \ll u$, in which case \Eq{eq:vlin} gives $J \approx \hat{J}$. [In other words, one may assume $F(J) \approx \delta(J - \hat{J})$.] From \Eq{eq:glargej}, one thereby obtains $G = a/2$, and thus \Eq{eq:drpw} predicts $\omega^2 = \omega_p^2$, as expected. 

Below we will calculate $\omega^2$ also for other representative cases, by formally constructing various asymptotic expansions of the integral in \Eq{eq:drpw}. Remarkably, \Eq{eq:gasym} will be sufficient for that; \ie knowing the exact form of $G(j)$ is not needed, except for evaluating insignificant numerical coefficients (\App{app:func}).

%%%%%%%%%%%%%%%%%%%%%%%%%%%%%%%%%%%%%%%%%%%%%%%%%%%%%%%%%%%%
\section{Smooth distributions} 
\label{sec:smooth}

First, let us consider the case when the distribution function $F(J)$ is smooth, \eg thermal. Of course, the wave \textit{dynamics} could not be described then within the present approach, on the score of particle unavoidable trapping and detrapping that would be associated with variations of $a$, rendering the wave nonadiabatic. Yet, assuming that the wave is quasihomogeneous and quasistationary (which are our necessary requirements in any case), its local dispersion must be approximately the same as that of a truly homogeneous and stationary wave \cite{foot:dissip}. Since the latter has $n^{(t)}$ fixed and is clearly adiabatic, our formalism is hence applicable for deriving the local NDR and can be built on as follows.

\begin{figure}
\centering
\includegraphics[width=.48\textwidth]{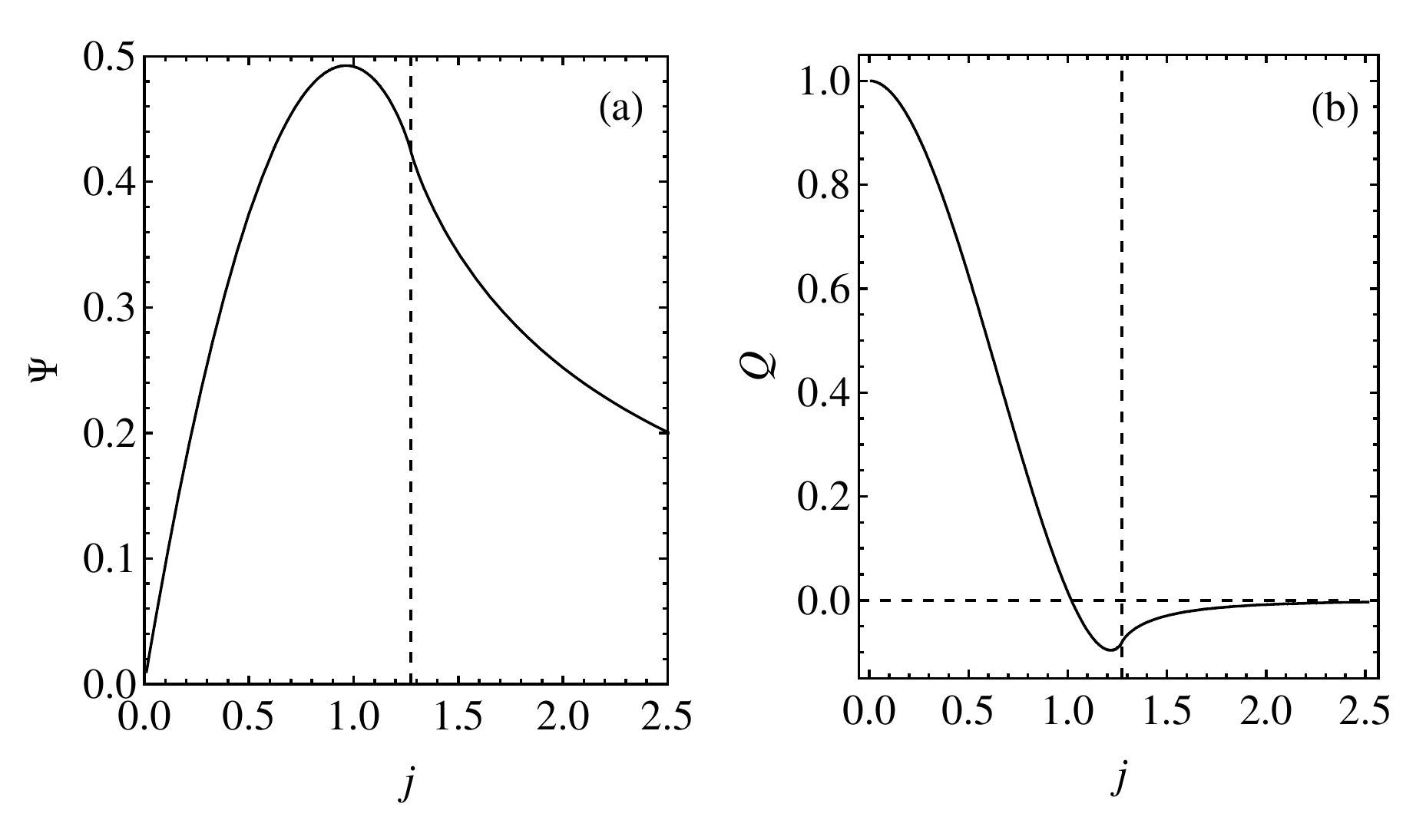}
\caption{(a)~Function $\Psi(j)$, \Eq{eq:psij}. (b)~Function $Q(j)$, \Eq{eq:qj}. The dashed lines show $j=j_*$ and also the asymptote of $Q(j)$. For details see \App{app:func}.}
\label{fig:pqj}
\end{figure}

Suppose that $F(J)$ remains finite at small $J$ or, at least, diverges less rapidly than $J^{-1}$. Then, one can take the integral in \Eq{eq:drpw} by parts and obtain
\begin{gather}\label{eq:drpw2}
1 - \frac{2\omega^2_p}{a\omega^2} \int^\infty_0 \Upsilon(J, a) F'(J)\,dJ = 0,
\end{gather}
where $\Upsilon(J, a) = \Psi(j)\hat{J}\sqrt{a}$, with the dimensionless function $\Psi(j)$ [\Fig{fig:pqj}(a)] introduced as
\begin{gather}\label{eq:psij}
\Psi(j) = - \int^j_0 G(\jmath)\,d\jmath.
\end{gather}
At $a \ll 1$, $\Upsilon$ changes rapidly with $J$ compared to $F(J)$. Then, without using the explicit form of $\Upsilon (J, a)$ but rather drawing on the leading terms in
\begin{gather}\label{eq:psiasym}
\Upsilon (J, a) = \left\{ 
\begin{array}{ll}
\ds J - \frac{J^2}{2\hat{J}a^{1/2}} + \ldots, & \quad J \ll \hat{J}a^{1/2},\\[10pt]
\ds \frac{a \hat{J}^2}{2J} + \frac{a^3 \hat{J}^6}{16 J^5}+\ldots, & \quad J \gg \hat{J}a^{1/2}
\end{array} 
\right.
\end{gather}
[obtained from \Eq{eq:gasym}], one can put \Eq{eq:drpw2} in an asymptotic form, specifically as follows. 

%-----------------------------------------------------------
\subsection{Asymptotic representation}

To approximate the integral
\begin{gather}\label{eq:U}
U = \frac{2}{a}\int^\infty_0 \Upsilon(J, a) F'(J)\,dJ,
\end{gather}
let us replace $\Upsilon F'$ with
\begin{gather}
\Upsilon F' \approx (\Upsilon F')_1 + (\Upsilon F')_2 - (\Upsilon F')_3.
\end{gather}
Here, the individual terms on the right-hand side correspond to asymptotics of $\Upsilon F'$ at small, large, and intermediate $J$, correspondingly, so that $\Upsilon F'$ is approximated accurately at all $J$. Specifically, we take
\begin{gather}\notag
(\Upsilon F')_i = \left\{ 
\begin{array}{l@{\quad}l}
\ds \Upsilon(J) \big[F'(0) + F''(0)J\big], & i = 1,\medskip \\ 
\ds \big[a\hat{J}^2/(2J)\big]\, F(J), & i = 2,\medskip \\
\ds \big[a\hat{J}^2/(2J)\big]\,\big[F'(0) + F''(0)J\big], & i = 3.
\end{array} 
\right.
\end{gather}
[Due to $\Upsilon - a\hat{J}^2/(2J) = \mc{O}(J^{-5})$ at large $J$, higher-order corrections are insignificant compared to those discussed below.] Hence, $U = U_1F''(0) + U_2$, where
\begin{gather}
U_1 = \frac{2}{a}\int^{\infty}_0 \bigg\{\Upsilon(J, a)\, J - \frac{a\hat{J}^2}{2}\bigg\}\, dJ, \notag \\[10pt]
U_2 = \frac{2}{a}\int^{\infty}_0 \bigg\{\Upsilon(J,a) F'(0) - \frac{a \hat{J}^2}{2J}\,\big[F'(J) - F'(0)\big] \bigg\}\, dJ.\notag
\end{gather}

It is convenient to rewrite $U_1$ in terms of the function
\begin{gather}\label{eq:qj}
Q(j) = 1 - 2j\Psi(j)
\end{gather}
[\Fig{fig:pqj}(b)], which we will also use below. Namely,
\begin{gather}
U_1 = - \varkappa\,\hat{J}^3 \sqrt{a}, \quad \varkappa = \int^\infty_0 Q(z)\,dz.
\end{gather}
The value of the numerical constant $\varkappa$ can be readily estimated from \Fig{fig:pqj}(b) as $\varkappa \sim 0.5$; for a precise calculation see \App{app:func}.

To simplify the expression for $U_2$, notice that
\begin{gather}\label{eq:aux1}
\frac{\pd U_2}{\pd a} = 2F'(0) \int^{\infty}_0 \frac{\pd}{\pd a}\left[\frac{\Upsilon(J,a)}{a}\right]dJ,
\end{gather}
which can be checked directly. On the other hand,
\begin{multline}
\frac{\pd}{\pd a}\left[\frac{\Upsilon(J,a)}{a}\right] 
 = \frac{\hat{J}^2}{J}\,\frac{d(\Psi j)}{dj}\,\frac{\pd j(J,a)}{\pd a} \\
 = -\frac{\hat{J}^2}{2a}\,\frac{d(\Psi j)}{dj}\,\frac{\pd j(J,a)}{\pd J}
 = - \frac{\pd}{\pd J}\left[\frac{\Upsilon(J,a) J}{2a^2}\right],
\end{multline}
and also
\begin{gather}
\int^{\infty}_0 \frac{\pd}{\pd J}\left[\frac{\Upsilon(J, a) J}{2a^2}\right]\,dJ = 
\frac{\Upsilon(J, a) J}{2a^2}\bigg|^\infty_0 = \frac{\hat{J}^2}{4a}.
\end{gather}
This yields
\begin{gather}
\pd_a U_2 = -\hat{J}^2F'(0)/(2a).
\end{gather}
Integrating the latter with respect to $a$ gives
\begin{gather}
U_2 = - \big[\hat{J}^2 F'(0)/2\big] \ln a + U_c.
\end{gather}
The integration constant $U_c$, independent of $a$, equals $U_2 (a = 1)$, so it can be written as
\begin{gather}
U_c = \int^{\infty}_0 \bigg\{2\Upsilon(J, 1) F'(0) + \frac{\hat{J}^2}{J}\,\big[F'(J) - F'(0)\big] \bigg\}\, dJ.\notag
\end{gather}
Using that $\Upsilon(J, 1) = \hat{J} \Psi(J/\hat{J})$, one gets
\begin{gather}\label{eq:uc}
U_c = \hat{J}^2 \int^{\infty}_0 \left[F'(J) - F'(0)\,Q\!\left(\frac{J}{\hat{J}}\right)\right]\,\frac{dJ}{J}.
\end{gather}
[Notice that, although determined by essentially nonlinear dynamics in the narrow vicinity of the resonance, $Q(J/\hat{J})$ nevertheless can affect the integrand on the thermal scale.] Finally, \Eq{eq:U} is summarized as
\begin{gather}\label{eq:uf}
U = - \big[\hat{J}^2 F'(0)/2\big] \ln a - \varkappa \sqrt{a}\,\hat{J}^3F''(0) + U_c.
\end{gather}

%-----------------------------------------------------------
\subsection{Nonlinear frequency shift}

Using \Eqs{eq:uc} and \eq{eq:uf}, we can rewrite \Eq{eq:drpw2} as
\begin{gather}\label{eq:ndr}
\epsilon(\omega, k)  + \frac{\omega_p^2}{2k^2}\,C_1 \ln a + \frac{\omega\omega_p^2}{k^3}\,\varkappa C_2\sqrt{a} = 0.
\end{gather}
Here we introduced
\begin{gather}\label{eq:epsilonF}
\epsilon = 1 -\frac{m^2\omega_p^2}{k^6} \int^{\infty}_0 \left[F'(J) - F'(0)\,Q\!\left(\frac{J}{\hat{J}}\right)\right]\,\frac{dJ}{J}
\end{gather}
and also
\begin{gather}\label{eq:cc}
C_1 = (m/k)^2F'(0), \quad C_2 = (m/k)^3F''(0).
\end{gather}
In particular, when $F'(0) = 0$, the nonlinear part of \Eq{eq:ndr} is small, yielding that the nonlinear frequency shift $\omega_{\rm NL}$ is also small; hence, 
\begin{gather}\label{eq:dwlin}
\omega_{\rm NL} = - \left(\frac{\pd \epsilon}{\pd \omega}\right)^{\!\!-1}\!\frac{\varkappa\omega_p^2}{k^2}\sqrt{\frac{e\tilde{\varphi}_0}{m}}\,C_2.
\end{gather}
Yet, at nonzero $F'(0)$, the nonlinear part of \Eq{eq:ndr} diverges logarithmically at small $a$; \ie wave interaction with resonant particles has a strong effect on $\omega$.

%-----------------------------------------------------------
\subsection{Comparison with existing theories}
\label{sec:comparison}

Equations \eq{eq:ndr}-\eq{eq:dwlin} generalize the existing NDR for quasimonochromatic eigenwaves in plasmas with smooth distributions of particles \cite{ref:dewar72b, ref:manheimer71a, ref:morales72, ref:lee72, ref:kim76, ref:barnes04, ref:rose01, ref:rose05, ref:lindberg07}, namely, as follows. First of all, notice that $\epsilon$, serving as a \textit{generalized linear dielectric function}, is a functional of the action distribution. Unlike the commonly used distribution of ``unperturbed'' velocities $f_0(v)$, which depends on how the wave was excited \cite{ref:dewar72b}, $F(J)$ is defined unambiguously; thus, the above equations hold for any excitation scenario. (Notice that finding $F(J)$ itself is kept as a separate problem, which is also where the effect of the quasistatic field \cite{ref:lindberg07} comes in.) Second, even if put in terms of $f_0(v)$, \Eqs{eq:ndr}-\eq{eq:dwlin} cover a wider class of particle distributions. The latter is seen as follows.

For example, consider a homogeneous wave developed with fixed $u$ slowly from zero amplitude, so each $J$ is conserved \cite{foot:P}, even through trapping and untrapping \cite{ref:best68, ref:timofeev78, ref:cary86, ref:lindberg07}. Then $F(J) = F_0(J)$, index 0 henceforth denoting the initial state. Yet, since there was no wave in that state, \Eq{eq:vlin} applies, so each $\ell$th derivative of $F_0(J)$ reads as
\begin{gather}
 F^{(\ell)}_0(J) = (k/m)^\ell \left[f_0^{(\ell)}(v^{+}) + (-1)^\ell f_0^{(\ell)}(v^{-})\right].
\end{gather}
Let us use bars to denote limits $f_0^{(\ell)}(v \to u \pm)$ and introduce
\begin{gather}
\bar{f}_0^{(\ell)}(v) = \left\{ 
\begin{array}{l@{\quad}l}
\ds \bar{f}^{(\ell)}_0(u-), & v < u,\medskip \\ 
\ds \bar{f}^{(\ell)}_0(u+), & v > u,
\end{array} 
\right.
\end{gather}
which is a piecewise-constant function equal to the left and right limits for $v < u$ and $v > u$ correspondingly; in particular,
\begin{gather}
C_1 = \bar{f}'_0(u+) - \bar{f}'_0(u-), \\ 
C_2 = \bar{f}''_0(u+) + \bar{f}''_0(u-).
\end{gather}
Hence $\epsilon$ rewrites as follows:
\begin{gather}\label{eq:eplin}
\epsilon = 1 -\frac{\omega_p^2}{k^2} \int^{\infty}_{-\infty} \,\frac{f_0'(v) - Q\bar{f}'_0(v)}{v-u}\,dv,
\end{gather}
where $Q \equiv Q(|v/u-1|)$. (Remarkably, contributing to $\epsilon$ are both passing and trapped particles.)

First, compare \Eq{eq:eplin} with the usual \cite{foot:epsilon}
\begin{gather}
\epsilon_{\rm L} = 1 - \frac{\omega_p^2}{k^2}\, \msf{P}\int^{\infty}_{-\infty} \frac{f_0'(v)}{v-u}\, dv,
\end{gather}
$\msf{P}$ denoting the principal value \cite{book:landau10}. For smooth $f_0(v)$, our $\epsilon$ can be put in the same form as $\epsilon_{\rm L}$, because $\msf{P}\int^{\infty}_{-\infty} (v-u)^{-1}Q\bar{f}'_0(v)\, dv = 0$. However, \Eq{eq:eplin} is valid also when $f'_0(v)$ is discontinuous across the resonance, a case in which $\epsilon_{\rm L}$ is undefined. This is because the integrand in \Eq{eq:eplin} is finite (piecewise-continuous), so the integral converges absolutely rather than existing only as a principal value (like $\epsilon_{\rm L}$ does). Second, for smooth $f_0(v)$, when $C_1 = 0$ and $C_2 = 2f''_0(u)$, \Eq{eq:dwlin} for $\omega_{\rm NL}$ matches that in \Ref{ref:dewar72b}, including the coefficient, since $\varkappa \approx 0.544$ (\App{app:func}). Yet, unlike the existing theory, our results apply just as well for arbitrary $C_1$ and $C_2$, in which case $f_0(v)$ may not be smooth while $F(J)$~is.

%%%%%%%%%%%%%%%%%%%%%%%%%%%%%%%%%%%%%%%%%%%%%%%%%%%%%%%%%%%%
\section{Beam-like distributions} 
\label{sec:beam}

Suppose now that, in addition to a smooth distribution, near the resonance there is a phase-space clump or a hole, further termed uniformly as a beam with ${F_b(J) \gtrless 0}$ and some average spatial density ${n_b \gtrless 0}$. Since particles contribute additively to the right-hand side in \Eq{eq:drpw}, we immediately obtain
\begin{gather}\label{eq:beammain}
\epsilon(\omega, k)  + \frac{\omega_p^2}{2k^2}\,C_1 \ln a + \frac{\omega\omega_p^2}{k^3}\,\varkappa C_2\sqrt{a} + \mcc{B} = 0,
\end{gather}
where $\mcc{B}$ is the beam-driven term,
\begin{gather}
\mcc{B} = - \frac{2\omega_b^2}{an_b\omega^2}\,\int^\infty_0 G F_b(J)\,dJ,
\end{gather}
with $\omega_b^2 = 4\pi n_be^2/m$. [Unlike $F(J)$, the distribution $F_b(J)$ is normalized to the beam density here.] For example, consider a beam of deeply trapped particles, so that $F_b(J) = n_b\delta(J)$. Since $G(0) = - 1$, one then has
\begin{gather}
\mcc{B} = 2\omega_b^2/(a\omega^2).
\end{gather}
Hence, when the beam nonlinearity dominates,
\begin{gather}
\epsilon(\omega, k) + 2\omega_b^2/\omega^2_E = 0.
\end{gather}
For instance, within the model \cite{my:dense}
\begin{gather}\label{eq:epsL}
\epsilon(\omega, k) = 1 - \frac{\omega_p^2}{\omega^2 - 3k^2 v_T^2},
\end{gather}
where $v_T$ is the thermal speed, one thereby gets the following dispersion relation:
\begin{gather}\label{eq:kras}
\omega^2 = \frac{\omega_p^2}{1 + 2\omega_b^2/\omega_E^2} + 3k^2 v_T^2,
\end{gather}
in agreement with \Ref{ref:krasovsky94}. Strictly speaking, the sinusoidal-wave approximation required for this result to hold applies only at $\omega_b^2/\omega_E^2 \ll 1$ \cite{ref:krasovsky94}, so one can further approximate \Eq{eq:kras} as
\begin{gather}
\omega^2 = \omega_p^2(1 - 2\omega_b^2/\omega_E^2) + 3k^2 v_T^2,
\end{gather}
in agreement with \Ref{ref:goldman71}. Since \Eq{eq:epsL} implies $kv_T \ll \omega$, one has $\omega \approx \omega_p$, so the above result also rewrites as
\begin{gather}
\omega^2 = \omega_{\rm L}^2 - 2\omega_b^2/a,
\end{gather}
where $\omega_{\rm L}^2 = \omega_p^2 + 3k^2 v_T^2$ corresponds to the linear limit. In particular, notice that $\omega_{\rm NL} = \mc{O}(a^{-1})$, with $\omega_{\rm NL} < 0$ for a clump and $\omega_{\rm NL} > 0$ for a hole. (Remember that $\omega_b^2 \propto n_b$ may have either sign.)

Now let $n_b$ itself depend on the wave amplitude. For example, a Van Kampen mode would have $\omega_b^2/a$ independent of $a$, yielding that trapped particles result in a \textit{linear} frequency shift $\omega_{\rm NL}$, in agreement with the original linear theory \cite{ref:vankampen55}. Also, let us consider the case when $F_b$ is constant across the trapping width, \ie $F_b(J) = \bar{F}_b\Theta(J) \Theta(J_* - J)$, with $\Theta$ being the Heaviside step function. Hence, $n_b = \bar{F}_bJ_*$, and 
\begin{gather}
\mcc{B} =  \frac{2\omega_b^2}{a\omega^2}\,\frac{\Psi(j_*)}{j_*} = \frac{8\bar{F}_b \hat{J}}{3\pi n\sqrt{a}}\,\frac{\omega_p^2}{\omega^2},
\end{gather}
where we used $\Psi(j_*) = 4/(3\pi)$ (\App{app:func}). Hence, the approximate formula is obtained, like above:
\begin{gather}\label{eq:aux2}
\omega^2 = \omega_{\rm L}^2 - \frac{8\omega_p^2\hat{J}F_b}{3\pi\sqrt{a} n}.
\end{gather}
Equation \eq{eq:aux2} also matches the result found previously, \eg in \Ref{ref:krasovsky07}.

%%%%%%%%%%%%%%%%%%%%%%%%%%%%%%%%%%%%%%%%%%%%%%%%%%%%%%%%%%%%
\section{Discussion} 
\label{sec:discuss}

The above analysis shows that the power index $\beta$ in the scaling for the nonlinear frequency shift $\omega_{\rm NL} \propto a^{\beta}$ depends critically on the properties of $F(J)$ at small $J$. The commonly anticipated $\beta = 1/2$ is realized when $F(J)$ is smooth, an example being the case when the wave develops adiabatically from zero amplitude. However, perturbations to $F$ on the trapping-width scale $J_*$ result in additional nonlinearities which yield $\beta < 0$. In particular, this affects the sign of $\pd_a \omega_{\rm NL}$, which can be decisive for the wave stability (Paper~III). Even more generally, describing such nonlinear kinetic effects in a fluid-like manner \cite{ref:dewar72d, ref:rose05, ref:rose08} is possible, in fact, only in special cases (\eg when $a(t)$ is monotonic and $u$ is fixed, as in \Ref{ref:yampolsky09a}), because $\omega_{\rm NL}$ is not a local function of $a$.

For example, consider a driven nonlinear mode with adiabatic chirp imposed \cite{ref:friedland06, ref:khain07, ref:khain10, ref:schmit10}. Due to the chirp, the wave phase velocity changes, carrying trapped particles along because of the autoresonance; then $F(J < J_*)$ remains intact. Yet $F(J > J_*)$ is changing, because passing particles do not conserve their actions as defined here \cite{foot:P}. Hence, even for an initially smooth $F(J)$, the chirp generally produces discontinuity of the action distribution at $J = J_*$, thereby changing the scaling for $\omega_{\rm NL}(a)$. Thus, accounting for nonlocality in $\omega_{\rm NL}(a)$ would be particularly important for trapped-particle waves whose phase velocity varies with time, modulated waves (Paper~III) being an example.

Finally, let us also consider effects that can be driven by dissipation, in case when plasma is collisional. Since $\omega$ changes rapidly with small $a$ in the presence of a phase-space clump or a hole, slow decay of $a$ will cause frequency downshifting or upshifting, correspondingly. Yet, since $\beta$ depends on how localized $F_b(J)$ is, another effect is anticipated, namely, as follows. Notice that a friction drag (say, proportional to the particle velocity) can cause condensation of the trapped distribution near the bottom of the wave potential trough \cite{my:mefffric}. Hence, peaking of $F(J)$ can occur, and $\beta$ can decrease gradually down to minus one. This represents a frequency sweeping mechanism additional to those considered in \Refs{ref:breizman10, ref:lilley10, ref:lilley09, ref:berk05}.

%%%%%%%%%%%%%%%%%%%%%%%%%%%%%%%%%%%%%%%%%%%%%%%%%%%%%%%%%%%%
\section{Conclusions}
\label{sec:concl}

In this paper, we expand on the results that were previously reported in our brief letter \cite{my:bgk}, performing a systematic study of kinetic nonlinearities in NDR of an adiabatic sinusoidal Langmuir wave in collisionless plasma. Specifically, a general NDR is derived in a nondifferential form, allowing for an arbitrary distribution of trapped electrons. The linear dielectric function is generalized, and the nonlinear kinetic frequency shift $\omega_{\rm NL}$ is found analytically as a function of the wave amplitude $a$. Smooth distributions yield $\omega_{\rm NL} \propto \sqrt{a}$, as usual. However, beam-like distributions of trapped electrons result in different power laws, or even a logarithmic nonlinearity, which are derived as asymptotic limits of the same dispersion relation. Such beams are formed whenever the phase velocity changes, because the trapped distribution is in autoresonance and thus evolves differently from the passing distribution. Hence, even adiabatic $\omega_{\rm NL}(a)$ is generally nonlocal.

%%%%%%%%%%%%%%%%%%%%%%%%%%%%%%%%%%%%%%%%%%%%%%%%%%%%%%%%%%%%
\section{Acknowledgments}

The work was supported through the NNSA SSAA Program through DOE Research Grant No. DE274-FG52-08NA28553.

\begin{figure}
\centering
\includegraphics[width=.48\textwidth]{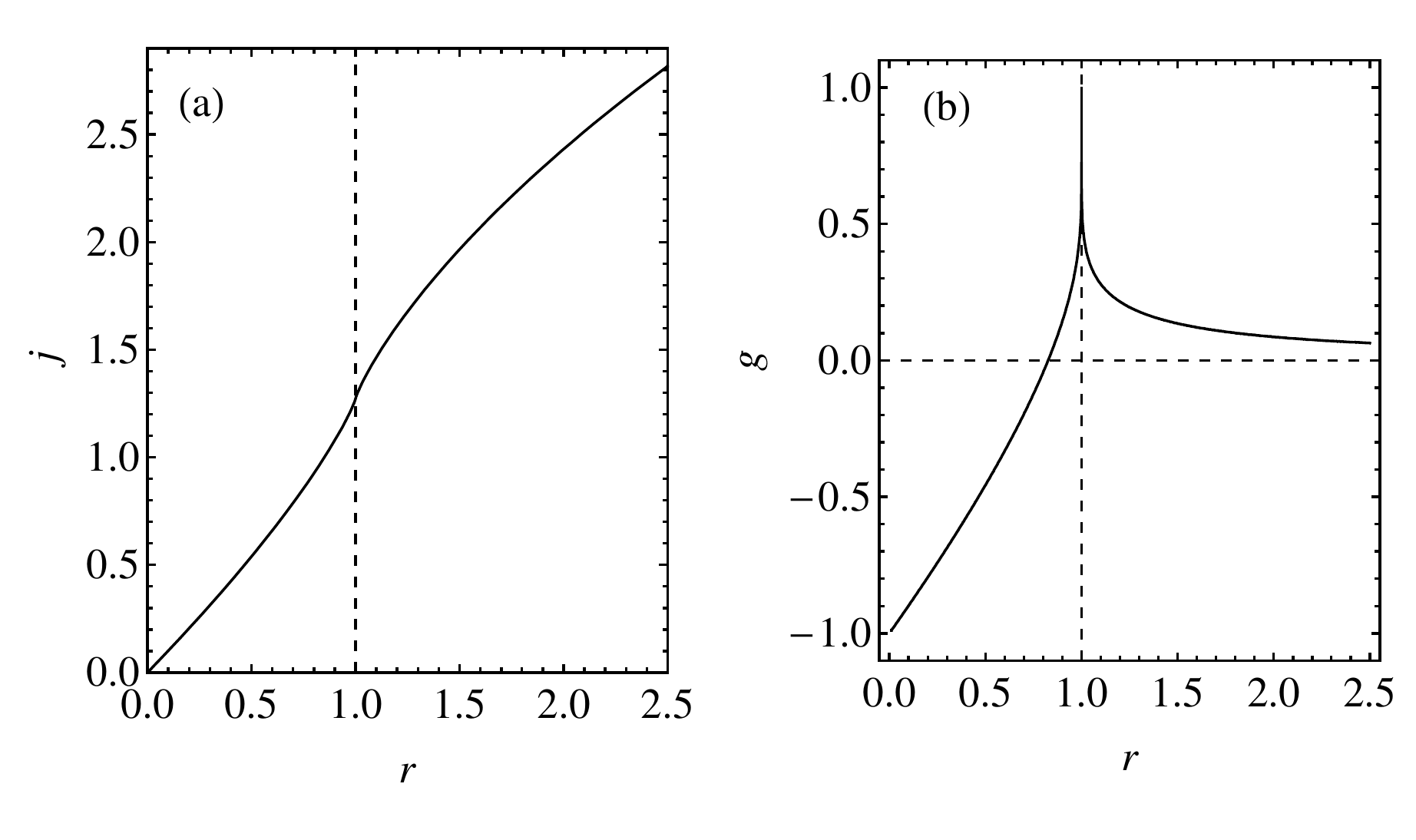}
\caption{(a)~Normalized action $j(r)$, \Eq{eq:jr}. (b)~Weight function $g(r)$, \Eq{eq:ge}. The dashed lines denote the separatrix $r = 1$ and also the asymptote of $g(r)$.}
\label{fig:jg}
\end{figure}

\newpage

\begin{figure}
\centering
\includegraphics[width=.48\textwidth]{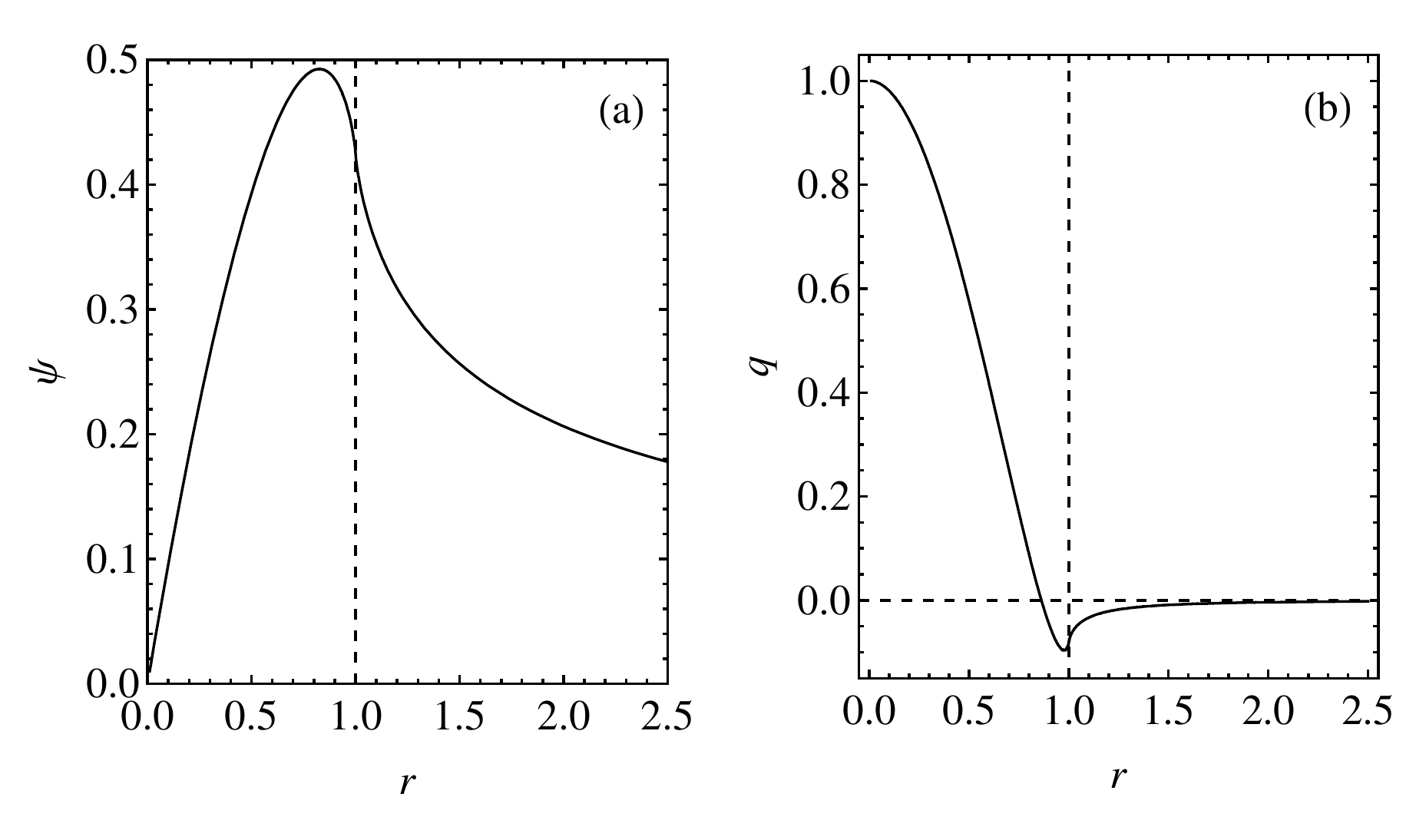}
\caption{(a)~Function $\psi(r)$, \Eq{eq:psir}. (b)~Function $q(r)$, \Eq{eq:qr}. The dashed lines denote the separatrix $r = 1$ and also the asymptote of $q(r)$.}
\label{fig:pqr}
\end{figure}

\appendix

%%%%%%%%%%%%%%%%%%%%%%%%%%%%%%%%%%%%%%%%%%%%%%%%%%%%%%%%%%%%
\section{Auxiliary functions}
\label{app:func}

Here, we summarize explicit expressions for the special functions used in this paper \cite{foot:math}. The analysis presented in the main text does not rely on these expressions \textit{per se}, yet those are needed to calculate the numerical values of $j_*$, $\psi(j_*)$, and $\varkappa$.

Specifically, the normalized action $j(r)$ reads as
\begin{gather}\label{eq:jr}
j(r) = \frac{4}{\pi}\times
\left\{ 
\begin{array}{ll}
\ds (r-1) \msf{K}(r) + \msf{E}(r), & \quad r < 1,\\[5pt]
\ds r^{1/2}\msf{E}(r^{-1}), & \quad r > 1
\end{array} 
\right.
\end{gather}
[\Fig{fig:jg}(a)], where $\msf{K}$ and $\msf{E}$ are the complete elliptic integrals of the first and second kind \cite{book:abramowitz},
\begin{gather}
\msf{K}(\zeta) = \int^{\pi/2}_0(1-\zeta\, \sin^2 \theta)^{-1/2}\,d\theta,\\
\msf{E}(\zeta) = \int^{\pi/2}_0(1-\zeta\, \sin^2 \theta)^{1/2}\,d\theta,
\end{gather}
with the following properties:
\begin{gather}
\msf{K}(0) = \msf{E}(0) = \pi/2, \quad \quad \msf{E}(1) = 1.
\end{gather}
In particular, one thereby gets $j(1) = 4/\pi$ (which can be found as well without introducing elliptic integrals), and
\begin{gather}\label{eq:jp}
j'(r) = \frac{2}{\pi} \times \left\{ 
\begin{array}{ll}
\ds \msf{K}(r), & \quad r < 1,\\[5pt]
\ds r^{-1/2}\msf{K}(r^{-1}), & \quad r > 1,
\end{array} 
\right.
\end{gather}
also yielding the oscillation period, with \Eq{eq:Omega}. From \Eq{eq:grgeneral}, one obtains then
\begin{gather}\label{eq:ge}
g(r) = \left\{ 
\begin{array}{ll}
\ds 1 - 2\msf{E}(r)/\msf{K}(r), & \quad r < 1,\\[5pt]
\ds 2r - 1 - 2r \msf{E}(r^{-1})/\msf{K}(r^{-1}), & \quad r > 1
\end{array} 
\right.
\end{gather}
[\Fig{fig:jg}(b)]. Further, the function $\Psi(j)$ is calculated as $\Psi(j) = \psi(r(j))$, with \Eq{eq:psij} yielding
\begin{gather}
\psi(r) = - \int^r_0 g(\bar{r})j'(\bar{r})\,d\bar{r}.
\end{gather}
Hence, \Eqs{eq:jp} and \eq{eq:ge} result in
\begin{widetext}
\begin{gather}\label{eq:psir}
\psi(r) = \frac{4}{3\pi}\times \left\{ 
\begin{array}{ll}
\ds (1 - r)\msf{K}(r)+ (2r - 1)\msf{E}(r), & \quad r < 1,\\[5pt]
\ds r^{1/2}[2(1-r)\msf{K}(r^{-1}) + (2r - 1)\msf{E}(r^{-1})], & \quad r > 1,
\end{array} 
\right.
\end{gather}
\end{widetext}
and $\psi(1) = 4/(3\pi)$ in particular [\Fig{fig:pqr}(a)].

Finally, $Q(j)$ is calculated as $Q(j) = q(r(j))$, with
\begin{gather}\label{eq:qr}
q(r) =1 - 2j(r)\psi(r)
\end{gather}
[\Fig{fig:pqr}(b)]. Hence,
\begin{gather}
\varkappa = \int^\infty_0 \left[1 - 2j(r)\psi(r)\right] j'(r)\,dr \approx 0.544.
\end{gather}

\vspace{-10pt}

%\bibliography{main,foot}

\begin{thebibliography}{10}

\bibitem{ref:whitham65}
G.~B. Whitham, J. Fluid Mech. {\bf 22}, 273 (1965).

\bibitem{book:whitham}
G.~B. Whitham, {\it Linear and Nonlinear Waves\/} (Wiley, New York, 1974), Chaps.~14 and 15.

\bibitem{tex:myacti}
I. Y. Dodin and N. J. Fisch, \textit{Adiabatic nonlinear waves with trapped
  particles: I. General formalism} (Paper~I), submitted together with the present paper.

\bibitem{ref:schamel00}
H.~Schamel, Phys. Plasmas {\bf 7}, 4831 (2000).

\bibitem{ref:krasovskii89}
V.~L. Krasovskii, Zh. Eksp. Teor. Fiz. {\bf 95}, 1951 (1989) [Sov. Phys. JETP
  {\bf 68}, 1129 (1989)].

\bibitem{ref:krasovskii95}
V.~L. Krasovskii, Zh. Eksp. Teor. Fiz. {\bf 107}, 741 (1995) [JETP {\bf 80},
  420 (1995)].

\bibitem{ref:benisti07}
D.~B\'enisti and L.~Gremillet, Phys. Plasmas {\bf 14}, 042304 (2007).

\bibitem{ref:matveev09}
A.~I. Matveev, Rus. Phys. J. {\bf 52}, 885 (2009).

\bibitem{ref:bohm49}
D.~Bohm and E.~P. Gross, Phys. Rev. {\bf 75}, 1851 (1949).

\bibitem{ref:khain07}
P.~Khain and L.~Friedland, Phys. Plasmas {\bf 14}, 082110 (2007).

\bibitem{my:bgk}
I.~Y. Dodin and N.~J. Fisch, Phys. Rev. Lett. {\bf 107}, 035005 (2011).

\bibitem{tex:myactiii}
I. Y. Dodin and N. J. Fisch, \textit{Adiabatic nonlinear waves with trapped
  particles: III. Wave dynamics} (Paper~III), submitted together with the present paper.

\bibitem{foot:anharm}
For example, for anharmonic oscillations, $\mc{E}(J, a)$ can
  be constructed iteratively, along the lines of existing iterative approaches
  to NDR \cite{ref:rose01, ref:lindberg07}. Yet, in practice, the
  spatial profile is most often assumed to be prescribed, \eg sinusoidal. Within
  this commonly accepted approach, our method becomes particularly useful,
  because then $\mc{E}(J, a)$ is known immediately.


\bibitem{ref:winjum07}
B.~J. Winjum, J.~Fahlen, and W.~B. Mori, Phys. Plasmas {\bf 14}, 102104 (2007).

\bibitem{ref:krasovsky94}
V.~L. Krasovsky, Phys. Scripta {\bf 49}, 489 (1994).

\bibitem{ref:rose01}
H.~A. Rose and D.~A. Russell, Phys. Plasmas {\bf 8}, 4784 (2001).

\bibitem{ref:krasovsky07}
V.~L. Krasovsky, J. Plasma Phys. {\bf 73}, 179 (2007).

\bibitem{ref:lindberg07}
R.~R. Lindberg, A.~E. Charman, and J.~S. Wurtele, Phys. Plasmas {\bf 14},
  122103 (2007).

\bibitem{foot:fluid}
Some of the fluid nonlinearities, yielding $\omega_{\rm NL} = \mc{O}(a^2)$, are retained in Eq.~(10) due to the nonlinear dependence $J(p)$. However, describing these effects accurately would require accounting for the contribution of the wave second harmonic, which is of the same order.

\bibitem{book:stix}
T.~H. Stix, {\it Waves in Plasmas\/} (AIP, New York, 1992), Sec.~8-6.

\bibitem{ref:dewar72b}
R.~L. Dewar, Phys. Fluids {\bf 15}, 712 (1972).

\bibitem{ref:ikezi78}
H.~Ikezi, K.~Schwarzenegger, and A.~L. Simons, Phys. Fluids {\bf 21}, 239
  (1978).

\bibitem{foot:amend}
To reduce the results of \Ref{ref:khain07} to our \Eq{eq:drpw}, the former must be amended in two aspects. First, rather than $n_i$, it is $f$ that must be adjusted to ensure that $\favr{n_e} = n_i$ (\ie $\favr{n_e}$ must be fixed); this changes $F_i$. Second, when summing over the contributions of multiple waterbags in Eqs.~(24)-(26) of \Ref{ref:khain07}, the weight must be $F(J)\,\Delta J$ rather than $f'\,\Delta u$. 

\bibitem{foot:dissip}
Nonzero $\pd_t a$ and $\pd_x a$ may also yield collisionless dissipation \cite{ref:benisti07, ref:benisti09, ref:benisti10, ref:ryutov73, ref:fahlen09, ref:fahlen11, ref:denavit72, ref:denavit75}, yet that would be weak within the conditions of geometrical optics and is not addressed~here.

\bibitem{ref:manheimer71a}
W.~M. Manheimer and R.~W. Flynn, Phys. Fluids {\bf 14}, 2393 (1971).

\bibitem{ref:morales72}
G.~J. Morales and T.~M. O'Neil, Phys. Rev. Lett. {\bf 28}, 417 (1972).

\bibitem{ref:lee72}
A.~Lee and G.~Pocobelli, Phys. Fluids {\bf 15}, 2351 (1972).

\bibitem{ref:kim76}
H.~Kim, Phys. Fluids {\bf 19}, 1362 (1976).

\bibitem{ref:barnes04}
D.~C. Barnes, Phys. Plasmas {\bf 11}, 903 (2004).

\bibitem{ref:rose05}
H.~A. Rose, Phys. Plasmas {\bf 12}, 012318 (2005).

\bibitem{foot:P}
In a homogeneous wave with varying $u$, passing particles conserve their
  canonical momentum [\Eq{eq:canP}] instead, as shown, \eg in
  \Refs{ref:lindberg07, ref:khain07}.

\bibitem{ref:best68}
R.~W.~B. Best, Physica 40 {\bf 40}, 182 (1968).

\bibitem{ref:timofeev78}
A.~V. Timofeev, Zh. Eksp. Teor. Fiz. {\bf 75}, 1303 (1978) [Sov. Phys. JETP
  {\bf 48}, 656 (1978)].

\bibitem{ref:cary86}
J.~R. Cary, D.~F. Escande, and J.~L. Tennyson, Phys. Rev. A {\bf 34}, 4256
  (1986).

\bibitem{foot:epsilon}
Remember that we consider quasistationary waves, on time scales large compared
  to the bounce period (Paper~I). No Landau damping exists in this case
  \cite{ref:mazitov65, ref:oneil65}, so the contribution from the pole in
  $\epsilon_{\rm L}$ is zero.

\bibitem{book:landau10}
E.~M. Lifshitz and L.~P. Pitaevskii, {\it Physical Kinetics\/} (Pergamon Press,
  New York, 1981), Sec.~29.

\bibitem{my:dense}
I.~Y. Dodin, V.~I. Geyko, and N.~J. Fisch, Phys. Plasmas {\bf 16}, 112101
  (2009).

\bibitem{ref:goldman71}
M.~V. Goldman and H.~L. Berk, Phys. Fluids {\bf 14}, 801 (1971).

\bibitem{ref:vankampen55}
N.~G.~Van Kampen, Physica {\bf 21}, 949 (1955).

\bibitem{ref:dewar72d}
R.~L. Dewar, W.~L. Kruer, and W.~M. Manheimer, Phys. Rev. Lett. {\bf 28}, 215
  (1972).

\bibitem{ref:rose08}
H.~A. Rose and L.~Yin, Phys. Plasmas {\bf 15}, 042311 (2008).

\bibitem{ref:yampolsky09a}
N.~A. Yampolsky and N.~J. Fisch, Phys. Plasmas {\bf 16}, 072104 (2009).

\bibitem{ref:friedland06}
L.~Friedland, P.~Khain, and A.~G. Shagalov, Phys. Rev. Lett. {\bf 96}, 225001
  (2006).

\bibitem{ref:khain10}
P.~Khain and L.~Friedland, Phys. Plasmas {\bf 17}, 102308 (2010).

\bibitem{ref:schmit10}
P.~F. Schmit and N.~J. Fisch, Phys. Plasmas {\bf 17}, 013105 (2010).

\bibitem{my:mefffric}
A.~I. Zhmoginov, I.~Y. Dodin, and N.~J. Fisch, Phys. Lett. A {\bf 375}, 1236
  (2011).

\bibitem{ref:breizman10}
B.~N. Breizman, Nucl. Fusion {\bf 50}, 084014 (2010).

\bibitem{ref:lilley10}
M.~K. Lilley, B.~N. Breizman, and S.~E. Sharapov, Phys. Plasmas {\bf 17},
  092305 (2010).

\bibitem{ref:lilley09}
M.~K. Lilley, B.~N. Breizman, and S.~E. Sharapov, Phys. Rev. Lett. {\bf 102},
  195003 (2009).

\bibitem{ref:berk05}
H.~L. Berk, Transp. Theory Stat. Phys. {\bf 34}, 205 (2005).

\bibitem{foot:math}
The results reported in the appendix were facilitated by \textsl{Mathematica}
  \textcopyright\ 1988-2009 Wolfram Research, Inc., version number 7.0.1.0.

\bibitem{book:abramowitz}
M.~Abramowitz and I.~A. Stegun, {\it Handbook of Mathematical Functions with
  Formulas, Graphs, and Mathematical Tables\/} (U.S. Dept. of Commerce: U.S.
  Gov. Printing Office, Washington, D.C., 1972), 10th~ed., vol.~55 of National
  Bureau of Standards Applied Mathematics Series, Sec.~17.3.

\bibitem{ref:benisti09}
D.~B\'enisti, D.~J. Strozzi, L.~Gremillet, and O.~Morice, Phys. Rev. Lett. {\bf 103}, 155002
  (2009).

\bibitem{ref:benisti10}
D.~B\'enisti, O.~Morice, L.~Gremillet, E.~Siminos, and D.~J. Strozzi, Phys.
  Plasmas {\bf 17}, 082301 (2010).

\bibitem{ref:ryutov73}
D.~D. Ryutov and V.~N. Khudik, Zh. Teor. Eksp. Fiz. {\bf 64}, 1252 (1973) [Sov.
  Phys. JETP {\bf 37}, 637 (1973)].

\bibitem{ref:fahlen09}
J.~E. Fahlen, B.~J. Winjum, T.~Grismayer, and W.~B. Mori, Phys. Rev. Lett. {\bf
  102}, 245002 (2009).

\bibitem{ref:fahlen11}
J.~E. Fahlen, B.~J. Winjum, T.~Grismayer, and W.~B. Mori, Phys. Rev. E {\bf
  83}, 045401(R) (2011).

\bibitem{ref:denavit72}
J.~Denavit and R.~N. Sudan, Phys. Rev. Lett. {\bf 28}, 404 (1972).

\bibitem{ref:denavit75}
J.~Denavit and R.~N. Sudan, Phys. Fluids {\bf 18}, 1533 (1975).

\bibitem{ref:mazitov65}
R.~K. Mazitov, Zh. Priklad. Mekh. Tekh. Fiz. {\bf 1}, 27 (1965).

\bibitem{ref:oneil65}
T.~O'Neil, Phys. Fluids {\bf 8}, 2255 (1965).

\end{thebibliography}

\end{document}